\documentstyle [12pt] {article}
\def\O{{\cal O}_{\rm ev}}
\def\r{{\rm rd}}
\title{Super Elliptic Curves}
\author{Jeffrey M. Rabin \\ Department of Mathematics \\ University of
California
at San Diego \\ La Jolla, CA 92093 \\ jrabin@ucsd.edu}
\date{January 1993}
\begin{document}
\maketitle
\begin{abstract}
A detailed study is made of super elliptic curves, namely super Riemann
surfaces
of genus one considered as algebraic varieties, particularly their relation
with
their Picard groups.
This is the simplest setting in which to study the geometric
consequences of the fact that certain cohomology groups of super Riemann
surfaces are not freely generated modules.
The divisor theory of Rosly, Schwarz, and Voronov gives a map from a supertorus
to its Picard group, but this map is a projection, not an isomorphism as it is
for ordinary tori.
The geometric realization of the addition law on Pic via intersections of the
supertorus with superlines in projective space is described.
The isomorphisms of Pic with the Jacobian and the divisor class group are
verified.
All possible isogenies, or surjective holomorphic maps between supertori, are
determined and shown to induce homomorphisms of the Picard groups.
Finally, the solutions to the new super Kadomtsev--Petviashvili (super KP)
hierarchy of Mulase--Rabin which arise from super elliptic curves via the
Krichever construction are exhibited.
\end{abstract}

\section{Introduction}
The theory of elliptic curves \cite{koblitz,silverman} is not only a rich and
fascinating subject in its own right, but a confluence of several major
branches
of mathematics and a source of simple and explicitly computable examples in
each.
These include Riemann surfaces, algebraic groups, Abelian varieties,
divisor theory, Diophantine equations, mapping class groups, and automorphic
functions.
The simple modular properties of the torus are of particular
importance in conformal field theory owing to the sewing axioms, by virtue of
which modular invariance on the torus guarantees this invariance at higher
genus, and in the related theory of elliptic genera.

The study of super elliptic curves, meaning super Riemann surfaces of genus
one considered as algebraic varieties, was initiated in
\cite{levin,rabinfreund}
with the use of superelliptic functions and super theta functions to embed
supertori in projective superspace as the sets of zeros of explicit polynomial
equations, generalizing the Weierstrass equation for an elliptic curve.
Missing from this work was any discussion of the group law on a superelliptic
curve.
Associated to any Riemann surface is its Picard group or Jacobian, the group of
line bundles of degree zero on the surface under tensor product.
A torus is itself a group because it is isomorphic to its Jacobian via the
classical Abel map.
The situation for supertori is more complicated because the Abel map turns out
to be a projection rather than an isomorphism.
The proof of this fact and extensive discussion of its consequences for the
theory of superelliptic curves are the subjects of this paper.

We study specifically the supertorus with odd spin structure given informally
(a more precise definition follows) as the quotient $M={\bf C}^{1,1}/G$ of the
complex superplane with coordinates $(z,\theta)$ by the supertranslation group
$G$ generated by the transformations
\begin{eqnarray}
& T: & z \rightarrow z + 1, \;\;\; \theta \rightarrow \theta, \nonumber \\
& S: & z \rightarrow z + \tau + \theta \delta, \;\;\; \theta \rightarrow \theta
+ \delta. \label{group}
\end{eqnarray}
The odd spin structure is of interest precisely because of the presence of the
odd modular parameter $\delta$ in addition to the usual even one $\tau$
(with the modulus $\tau$ and the theta functions, we will tolerate some
exceptions to the standard convention that Greek letters denote odd quantities
while Roman letters denote even ones).
Meromorphic functions on the supertorus are just
meromorphic functions $F(z,\theta)$ on ${\bf C}^{1,1}$ which are $G$-invariant,
or superelliptic.
In particular, the
cohomology group $H^0(M,{\cal O})$ consisting of global holomorphic
functions is easily shown to be the set of functions  $a +\alpha\theta$ with
constant coefficients $a,\alpha$ such that $\alpha\delta = 0$.
For even functions, $a$ should be even and $\alpha$ odd.
Owing to the constraint on $\alpha$, this is {\it not} simply the
vector superspace ${\bf C}^{1,1}$ with basis $\{1,\theta\}$; it is indeed a
module over the Grassmann algebra $\Lambda$ containing all our
odd parameters, but this module is not freely generated.
This situation occurs generically for super Riemann surfaces with odd spin
structure \cite{giddings,problems} and its implications are not well understood
in general.
The primary motivation for this work was to study them in this simplest case,
in which complete, explicit calculations are possible and illuminating.

One is so accustomed to the fact that a sheaf cohomology group $H^i(M,{\cal
F})$
carries the structure of a finite-dimensional vector space that one forgets
that
the proof is nontrivial \cite{hartshorne}.
Certainly the existence of this structure is so central to geometric
applications of cohomology that one would hardly know where to begin without
it: the Riemann-Roch theorem is only the simplest of the tools designed to
compute the dimensions of these vector spaces.
Such tools only generalize in the super case for generic {\it even} spin
structures, the ``normal case'' considered in \cite{rsv}.
The lack of a super vector space structure causes difficulties in the theory
and applications of super Riemann surfaces whenever a basis for a
cohomology space of functions, differentials, or deformations would be
desirable.
(One should note that because $\Lambda$ is a Grassmann algebra over
$\bf C$, the cohomology groups do have vector space structures over {\bf C}.
However, because one may wish to vary the Grassmann algebra, it is the module
structure over $\Lambda$ that is of interest.)
Certainly the super Riemann-Roch theorem holds only in the normal case or under
additional assumptions.
In the application to superstrings, bases for spaces of holomorphic
differentials of various weights are normally used to express the
superdeterminants appearing in the path integral measure and in the expressions
for amplitudes used in finiteness and unitarity proofs.
These analyses are considerably more complicated when such free bases do not
exist \cite{dhoker}.
The geometry associated to the super KP hierarchies \cite{rabin,mulase} would
normally be described in terms of a super Grassmannian of vector subspaces of,
say, functions on the supercircle \cite{schwarz} with the Krichever map sending
a supercurve and additional geometric data to the vector subspace given by a
suitable cohomology group.
When the cohomology lacks a vector space structure, this construction fails.
Presumably the correct super Grassmannian contains certain $\Lambda$-submodules
as well as free subspaces, but the specific class of submodules and the
geometry
of the resulting Grassmannian have not been elucidated.
The operator formalism for fermionic string makes use of the same Grassmannian
structures \cite{nelson}, and the modifications which might be required here in
the non-normal case deserve investigation.
Sometimes the problem is finessed by considering split supercurves (no
supermoduli) and transporting the results to the rest of supermoduli space by
using the fermionic stress tensor as a connection.
Fully justifying this procedure would require an understanding of how the
stress
tensor encodes the structure of the submodules during transport through the
Grassmannian.

This paper concentrates on how the non-free character of the cohomology affects
the geometry of a superelliptic curve, particularly its relation to its
Jacobian.
Section 2 develops the basics of function theory.
We exhibit the building blocks for the explicit construction of functions, the
super analogues of Weierstrass $\wp$ functions and theta functions, as well
as deriving the general constraints on the divisor of a superelliptic function.
Because the canonical bundle of a superelliptic curve is trivial, this analysis
applies to meromorphic differentials of all weights as well as to functions.
In Section 3 we explicitly compute the Picard group (group of line bundles),
the Jacobian (space of linear functionals modulo periods), and the divisor
class group (divisors modulo divisors of functions) of a superelliptic curve,
verifying that they are all isomorphic.
This isomorphism has been proven for all super Riemann surfaces in the normal
case \cite{rsv}, but not more generally thus far.
The Abel map from the curve to its Jacobian is obtained and observed to be a
projection $\pi$: it takes the quotient of the curve by the relation
$(z,\theta) \sim (z + \alpha \delta,\theta)$ for all $\alpha$.
The origin of this extra identification is traced to the necessity of
abelianizing the nonabelian group $G$ in order for the quotient to admit a
group
structure.
Section 4 shows that, modulo this identification and an ambiguity in the choice
of identity element, the group operation on the Jacobian can be performed
geometrically on the curve by intersecting it with special planes in the
standard
superprojective embedding.
Section 5 determines all the isogenies of superelliptic curves.
These are surjective holomorphic mappings between supertori.
For elliptic curves one proves that they are necessarily
homomorphisms in the group structure.
Here, since a superelliptic curve does not carry the group structure of its
Jacobian, the best one can do is to show that an isogeny induces a homomorphism
of the Jacobians via the projection $\pi$.
We also study isogenies of a superelliptic curve to itself and show that a
nonsplit curve admits only trivial endomorphisms.
Section 6 contains a major application of these results to the new super KP
system discovered by Mulase and the author \cite{rabin,mulase}.
This system of nonlinear PDEs for the coefficients of a pseudosuperdifferential
operator describes, via the Krichever construction, the deformation of a line
bundle $\cal L$ over an algebraic supercurve by certain commuting flows in the
Jacobian.
The pseudodifferential operator is closely related to a special section of
$\cal L$ called the Baker-Akhiezer function.
The algebraic supercurves involved are generally not super Riemann surfaces
except in the special case of genus one.
In this exceptional case we can construct explicit solutions to the super KP
system describing flows in the Jacobian of a superelliptic curve, in terms of
Weierstrass elliptic functions.
The result can be presented as an isomorphism between a ring of meromorphic
functions on the superelliptic curve and a ring of supercommuting differential
operators \cite{rabin,supercomm}.
It generalizes the classical result that the operators
\begin{eqnarray}
Q & = & \frac{d^2}{dx^2} - 2 \wp(x+a), \\
P & = & Q_+^{3/2} = \frac{d^3}{dx^3} - 3 \wp(x+a)\frac{d}{dx} - \frac{3}{2}
\wp'(x+a)
\end{eqnarray}
arising from an elliptic curve generate a commutative ring.
The parameter $a$ should be viewed as a coordinate on the Jacobian and varies
linearly with the flow parameters.
A new feature of the super case is that the
supercommutativity of the ring depends upon the fact that the theta
function satisfies the heat equation.
Section 7 contains conclusions and directions for further research.
An Appendix briefly considers the problem of finding rational points on
superelliptic curves.
Here the nilpotent elements of $\Lambda$ linearize the problem to locating
rational points on the (co)tangent line to an elliptic curve at a rational
point,
so nothing of number-theoretic interest has been added.
Throughout this paper, computations which employ standard methods are
nevertheless given in considerable detail, so as to remove any mystery from the
supermodulus $\delta$ and display clearly the role it plays in modifying the
classical results.

Before proceeding, let us return to the precise definition of the superelliptic
curves we study.
We fix a finite-dimensional complex Grassmann (exterior) algebra $\Lambda$ in
which $\delta$ is an odd element and $\tau$ an even one with
${\rm Im}\; \tau_\r >0$.
(Throughout this paper the subscript ``rd'' on a Grassmann variable,
supermanifold, supergroup, etc.\ denotes the reduction of this object by
modding
out the ideal of nilpotents in $\Lambda$ or in the structure sheaf.)
We adopt the standard sheaf-theoretic
treatment of supermanifolds \cite{maninbook} within which we are really dealing
with families of superelliptic curves over the parameter superspace  ${\cal B}
=
({\rm pt},\Lambda)$.
Our covering space, informally denoted ${\bf C}^{1,1}$, is really the trivial
family ${\bf C}^{1,1} \times {\cal B}$, meaning the complex plane $\bf C$
equipped with the structure sheaf ${\cal O}_{\bf C} \otimes \Lambda[\theta]$,
where $\Lambda[\theta]$ is the larger Grassmann algebra whose generators are
$\theta$ and the generators of $\Lambda$.
The family of superelliptic curves $M$ over $\cal B$ is the quotient of this
family by the group $G$, meaning the following.
The reduced space of
$M$ is the standard torus $M_\r$ with modular parameter $\tau_\r$.
The structure sheaf of $M$ assigns to any open set $U$ of $M_\r$ the following
ring ${\cal O}_U$.
$U$ is covered by a collection of connected open sets $U_i$ of $\bf C$.
To each element $g$ in $G$ there corresponds a transformation
$g_\r$ in the reduced group $G_\r$ generated by
\begin{eqnarray}
& T_\r : & z \rightarrow z+1, \nonumber \\
& S_\r : & z \rightarrow z + \tau_\r,
\end{eqnarray}
which maps each $U_i$ to some (possibly the same) $U_j$.
For ${\cal O}_U$ we take all collections of functions
$\{F_i(z,\theta) \in {\cal O}_{U_i}\}$ which are $G$-invariant in the
sense that $F_j(z,\theta) = gF_i(z,\theta)$ whenever $U_j=g_\r U_i$, $g \in G$.
Here $g$ acts on functions via Taylor expansion in nilpotents as usual:
$F(z + \tau + \theta \delta,\theta + \delta)$ means
$F(z + \tau,\theta) + \theta \delta \partial_z F(z + \tau,\theta) +
\delta \partial_\theta F(z + \tau,\theta)$.
If $\tau$ has a nilpotent part then the
last expression is defined by further Taylor expansion in this nilpotent part.
The statement that $\rho: M \rightarrow {\cal B}$ is a family means that there
is
a pullback map of the functions $\Lambda$ on $\cal B$ to functions on $M$; the
elements of $\Lambda$ play the role of global constant functions on $M$ and as
such all the cohomology groups of $M$ are modules over $\Lambda$ (or its even
part if the sheaf is purely even or odd).

For those readers less comfortable with sheaf-theoretic language, which often
includes the author, we can consider the set of $\Lambda$-valued points of $M$
rather than $M$ itself.
This is the set of (even) maps
${\cal B} \rightarrow M \stackrel{\rho}{\rightarrow} {\cal B}$
for which the composed map ${\cal B} \rightarrow {\cal B}$ is the identity.
For each point of $M$ this is an evaluation of functions at that point by
assigning even and odd values from $\Lambda$ to the coordinates $z$ and
$\theta$
respectively.
That is, it is just an abstract description of the
$\Lambda$-supermanifolds of \cite{rsv}, or the supermanifolds of DeWitt
\cite{dewitt} or Rogers \cite{rogersmanifolds}, which are genuine sets of
points
with Grassmann-valued coordinates.
The Picard and Jacobian groups as defined here naturally appear as such sets of
$\Lambda$-valued points and will be discussed as such; our constructions can be
translated into pure sheaf-theoretic terms by those readers with the
sophistication to prefer this viewpoint.

The choice of Grassmann algebra will usually be left open, but two cases are
worth distinguishing.
One is the case in which $\delta$ is one of the generators of
$\Lambda$.
The most important example is the two-dimensional algebra having $\delta$ as
its
only generator (plus unity);
if we let $\tau$ run through the upper half-plane this gives the universal
Teichm\"{u}ller family of supertori (apart from the identification of $\pm
\delta$).
The other is the general case in which $\delta$ is an element of
$\Lambda$ but not necessarily a generator.
Such a family is a pullback of the
universal family by a map of the base spaces, which indeed pulls back $\delta$
to some element of $\Lambda$, e.g. $\delta = \beta_1\beta_2\beta_3$ in terms of
generators $\beta_i$.
The most important distinction between these cases is that
when $\delta$ is a generator it annihilates only multiples of itself, while in
general it may annihilate other elements as well, e.g. multiples of $\beta_1$
in
the above example.

\section{Basic Function Theory}

In order to construct explicit functions and sections of bundles on the
supertorus $M$, in particular the Baker-Akhiezer function appearing in super KP
theory, we need the building blocks corresponding to the Weierstrass elliptic
function $\wp(z;\tau)$ and the theta function $\Theta(z;\tau)$
(the capital letter is used for theta functions in this paper
to avoid confusion with the coordinate $\theta$) introduced in
\cite{rabinfreund}.

The super Weierstrass function is
\begin{equation}
R(z,\theta;\tau,\delta) = \wp(z;\tau + \theta\delta) = \wp(z;\tau) +
\theta \delta \dot{\wp}(z;\tau),
\end{equation}
where by convention a dot denotes $\partial_\tau$ while a prime will mean
$\partial_z$.
It is superelliptic, as are its supercovariant derivatives
$D^nR$, where $D = \partial_\theta + \theta \partial_z$ commutes with the
generators of $G$ and satisfies $D^2 = \partial_z$.
These functions provide the standard embedding of $M$ in projective superspace
which we will recall in Section 4.

Similarly, our super theta function will be
\begin{equation}
H(z,\theta;\tau,\delta) = \Theta(z; \tau + \theta \delta).
\end{equation}
The ordinary theta function appearing here is the one often denoted
$\Theta \left[ \begin{array}{c} \frac{1}{2} \\ \frac{1}{2} \end{array} \right]
(z;\tau)$, which corresponds to the odd spin structure.
It has a simple zero at $z=0$ and the other lattice points, and satisfies
\begin{eqnarray}
\Theta(z+1;\tau) & = & -\Theta(z;\tau) = \Theta(-z;\tau), \nonumber \\
\Theta(z+\tau;\tau) & = & -e^{-\pi i \tau - 2 \pi i z} \Theta(z;\tau).
\end{eqnarray}
As a result, the super theta function satisfies
\begin{eqnarray}
& & H(z+1,\theta) = -H(z,\theta) = H(-z,\theta), \nonumber \\
& & H(z + \tau + \theta \delta, \theta + \delta) =
-e^{- \pi i \tau - \pi i \theta \delta - 2 \pi i z} H(z,\theta),
\label{phases}
\end{eqnarray}
where the moduli dependence of $H$ has been suppressed.
The relation between $\Theta$ and $\wp$ is \cite{chandra}
\begin{equation}
\frac{d^2}{dz^2} \log \Theta(z;\tau) = -\wp(z;\tau) + q, \;\;\;\;
q = \frac{\Theta'''(0;\tau)}{3\Theta'(0;\tau)}. \label{q}
\end{equation}
The first derivative $\partial_z \log \Theta$ is nearly elliptic, being
invariant under $z \rightarrow z+1$ and changing by an additive constant under
$z \rightarrow z+\tau$.
Since this is also the behavior of $\theta$ according to (\ref{group}), we can
form the superelliptic combination \cite{nepo}
\begin{equation}
\sigma(z,\theta;\tau,\delta) = \theta + \frac{\delta}{2\pi i} \frac{d}{dz} \log
\Theta(z;\tau)   \label{sigma}
\end{equation}
which reduces to $\theta$ in the split case where $\delta=0$.
This function will be of particular importance in view of the fact that it is
holomorphic in the split case (when cohomology is freely generated) but only
meromorphic otherwise.

To describe the meromorphic functions on $M$ and construct them from the
building blocks above, we turn to the study of divisor theory.
In the usual Cartier divisor theory, a divisor would be a subvariety of
codimension $(1,0)$, hence dimension $(0,1)$, given locally by an even equation
$F(z,\theta)=0$.
The fact that such divisors are not points breaks the strong analogy between
elliptic and superelliptic curves.
It was the great insight of Rosly, Schwarz, and Voronov \cite{rsv} (see also
\cite{manin}) to make use of the covariant derivative $D$ (the superconformal
structure) which exists locally on any super Riemann surface to define divisors
of codimension $(1,1)$ --- points --- via the simultaneous solutions of
\begin{equation}
F(z,\theta) = 0, \;\;\;\; DF(z,\theta) = 0.
\end{equation}
For any even function $F$ for which the reduced function $F_\r(z)$ is not
identically zero, a point ($\Lambda$-valued!) $(z_0,\theta_0)$ satisfying these
equations is called a principal zero of $F$.
If we write $F(z,\theta) = f(z) + \theta\phi(z)$ and assume that $(z_0)_\r$ is
a
simple zero of $f_\r$ (in this case we are discussing a principal simple zero
of
$F$), this amounts to the statements
\begin{equation}
f(z_0) = 0, \;\;\;\; \theta_0 = -\phi(z_0)/f'(z_0).
\end{equation}
A principal pole of $F$ is a principal zero of $1/F$.
A formal sum of points $\sum n_iP_i$ is a divisor of $F$ provided that in a
chart containing  $P_i = (z_i,\theta_i)$ we can write
\begin{equation}
F(z,\theta) = E(z,\theta) \prod_i (z - z_i - \theta \theta_i)^{n_i},
\end{equation}
where the product is over the $P_i$ contained in the chart and $E$ is
holomorphic with $E_\r \neq 0$ in this chart [it may not be possible to
separate
all the points $P_i$ because the corresponding reduced points $(z_i)_\r$ may
coincide].
A subtlety is that a single function may have more than one divisor if its
zeros and poles are not simple.
For example, on ${\bf C}^{1,1}$, $F = (z+a)^2 = z(z+2a)$ with nilpotent even
constant $a$ satisfying $a^2=0$ has the two distinct divisors of zeros
$2(-a,0)$
and $(0,0)+(-2a,0)$ as well as others.
On the supertorus, $R(z,\theta)$ has a principal double pole at $(0,0)$ and two
simple zeros.
The super theta function $H(z,\theta)$, actually a section of a
bundle rather than a function, has a principal simple zero at $(0,0)$.

We now derive the necessary and sufficient condition for a divisor
$\sum n_iP_i$ to be a divisor of some meromorphic function $F$ on $M$: the sum
of
the $P_i$ with multiplicity must differ from a lattice point by
$(\alpha \delta,0)$ for some constant $\alpha$, namely
\begin{eqnarray}
\sum_i n_i \theta_i & = & n \delta, \nonumber \\
\sum_i n_i z_i & = & m + n \tau + \alpha \delta, \label{sumrules}
\end{eqnarray}
for integers $m,n$.
Of course, the total degree $\sum_i n_i$ must also vanish because it vanishes
for the divisor of the reduced function on the torus $M_\r$.

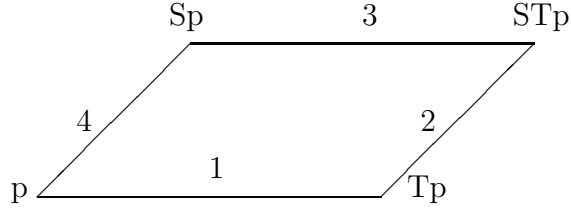
\begin{figure}
\begin{center}
\begin{picture}(220,80)
\put (20,0){\line(1,0){130}}
\put (150,0){\line(1,1){58}}
\put (208,58){\line(-1,0){130}}
\put (78,58){\line(-1,-1){58}}
\put (10,0){p}
\put (160,0){Tp}
\put (200,65){STp}
\put (70,65){Sp}
\put (85,7){1}
\put (165,25){2}
\put (143,65){3}
\put (35,25){4}
\end{picture}
\caption{The period parallelogram, an integration contour for the proof of sum
rules for the divisor of a superelliptic function. Except for orientation,
sides 1 and 3 are related by the supertranslation S, sides 2 and 4 by T.}
\end{center}
\end{figure}

The proof of the necessity follows the classical and elementary proof for
elliptic curves \cite{koblitz} by integrating $DF/F = D \log F$ around a period
parallelogram as shown in Fig. 1,
chosen to avoid the points of the divisor.
An easy computation shows that near a principal pole or zero where $F$ behaves
as $(z-z_i-\theta\theta_i)^{n_i}$, we have \begin{equation}
\frac{DF}{F} \sim \frac{n_i(\theta-\theta_i)}{z-z_i-\theta\theta_i} =
\frac{n_i(\theta-\theta_i)}{z-z_i},
\end{equation}
plus holomorphic terms.
Then we evaluate the following two contour integrals (For details on the
definition of super contour integration, see \cite{friedan,mcarthur,rogers}.
For closed contours it is simply Berezin integration over $\theta$ followed by
ordinary contour integration.
For an open contour lying in a simply connected region in which $F$ is
holomorphic, it is the change in an antiderivative $\Phi$, with $D\Phi=F$,
between the endpoints.):
\begin{eqnarray}
\oint \theta \frac{DF}{F} {\bf dz} & = &
\sum_i \oint \frac{-n_i\theta\theta_i}{z-z_i} dz \,d\theta  \nonumber \\
& = & 2 \pi i \sum_i n_i \theta_i,
\end{eqnarray}
and similarly
\begin{eqnarray}
\oint z \frac{DF}{F} {\bf dz} & = & \sum_i \oint \frac{n_iz\theta}{z-z_i} dz \,
d\theta  \nonumber  \\
& = & \sum_i n_i \oint \left( 1 + \frac{z_i}{z-z_i} \right) dz  \nonumber  \\
& = & 2 \pi i \sum_i n_i z_i.
\end{eqnarray}

Next we evaluate the integrals over each side of the parallelogram and
use the fact that $F$ is the same on opposite sides by superellipticity.
For the first integral we note that $\theta$ is the same on sides 2 and 4,
which
have opposite orientations, so those contributions cancel, while sides 1 and 3
are related by $\theta \rightarrow \theta + \delta$.
The Jacobian factors relating these integrals are unity, which is also clear
from the antiderivative definition and the fact that $D$ commutes with the
generators of $G$.
Hence these contributions sum to
\begin{eqnarray} \oint \theta \frac{DF}{F} {\bf dz} & = & \int_1 -\delta
\frac{DF}{F} {\bf dz} = -\delta \int_1 D \log F {\bf dz}  \nonumber \\
& = & 2 \pi i n \delta,
\end{eqnarray}
the point being that only the reduced part
of $\log F$ is multivalued, the nilpotent part involving derivatives of $\log$
via the Taylor expansion.
Comparing with the previous evaluation of the integral
gives the sum rule for $\theta_i$.
For the $z$ integral things are slightly more
complicated. Sides 1 and 3 are related by $z \rightarrow z + \tau + \theta
\delta$, sides 2 and 4 by $z \rightarrow z + 1$.
Making these substitutions gives
\begin{eqnarray}
\oint z \frac{DF}{F} {\bf dz} & = & \int_1 (-\tau - \theta \delta) \frac{DF}{F}
{\bf dz} + \int_2 \frac{DF}{F} {\bf dz}  \nonumber  \\
& = & 2 \pi i (m + n\tau) + \delta \int_1 \theta \frac{DF}{F} {\bf dz},
\end{eqnarray}
where the last integral can have any odd value.
Calling it $-2\pi i\alpha$, we obtain the sum rule for $z_i$.

To show the sufficiency, we construct a function having any given divisor
satisfying the sum rules in terms of the super theta function.
First we note the effect of a supertranslation on the divisor of a function:
if $F(z,\theta)$ has the behavior $(z-z_i-\theta\theta_i)^{n_i}$ corresponding
to a principal zero or pole at $(z_i,\theta_i)$, then
\begin{equation}
F(z-a-\theta\epsilon,\theta-\epsilon) \sim [z-(z_i+a+\theta_i\epsilon) -
\theta(\theta_i+\epsilon)]^{n_i},  \label{shift}
\end{equation}
shifting the zero or pole to $(z_i+a+\theta_i\epsilon,\theta_i+\epsilon)$.
The odd coordinates of the divisor are shifted uniformly by $\epsilon$, the
even coordinates uniformly by $a$ but also nonuniformly by a term proportional
to the odd coordinates.
This changes the sum of the $z_i$ by a multiple of the sum of the $\theta_i$,
which is a multiple of $\delta$, consistent with the sum rule for $z_i$.
In particular, the theta function $H(z-z_i-\theta\theta_i,\theta-\theta_i)$ is
holomorphic with a principal simple zero at $(z_i,\theta_i)$.

Unfortunately, this theta function is not convenient for our purposes since it
does not transform by a mere phase under the group $G$.
As a consequence of the commutation relations of supertranslations, the
generator $S$ sends it to a phase times
$H(z-z_i-\theta\theta_i-2\delta\theta_i,\theta-\theta_i)$.
However, the function $H(z-z_i-\theta\theta_i,\theta+\theta_i)$ also has a
principal simple zero at $(z_i,\theta_i)$ and transforms as
\begin{equation}
SH(z-z_i-\theta\theta_i,\theta+\theta_i) =
-e^{-\pi i [\tau + (\theta+\theta_i)\delta + 2(z-z_i-\theta\theta_i)]}
H(z-z_i-\theta\theta_i,\theta+\theta_i).   \label{Sphase}
\end{equation}
This remedy of changing the relative sign in $\theta-\theta_i$ amounts to the
usual replacement of a SUSY generator by a SUSY covariant derivative.

Let us suppose first that $\sum_i n_iP_i$ is a degree-zero divisor for which
the $P_i$ sum exactly to a lattice point, with no remainder $\alpha\delta$.
By adding the fictitious points $(0,0)-(m+n\tau,n\delta)$ we can assume that
the
$P_i$ sum to zero without changing the divisor on $M$.
Then a superelliptic function with this divisor is
\begin{equation}
F(z,\theta) = \prod_i [H(z-z_i-\theta\theta_i,\theta+\theta_i)]^{n_i}.
\label{thetas}
\end{equation}
Its invariance under the generators of the group $G$ is easily checked using
the relation (\ref{Sphase}) and the sum rules (\ref{sumrules}).

The simplest example of a degree-zero divisor satisfying the sum rules with a
nontrivial remainder $\alpha\delta$ is $\Delta = (\alpha\delta,0)-(0,0)$.
A meromorphic function with this divisor is easily constructed from the
function $\sigma$ introduced in Eq. (\ref{sigma}), namely
\begin{equation}
F_\Delta(z,\theta) = 1 - 2 \pi i \alpha \sigma(z,\theta) =
(1 - 2 \pi i \alpha \theta)[1 - \alpha \delta \frac{d}{dz} \log
\Theta(z;\tau)],
\end{equation}
where the second form shows the behavior $1 - \frac{\alpha\delta}{z}$ near
$z=0$ dictated by the divisor.
Now, given an arbitrary divisor satisfying the sum rules, subtracting the
divisor $\Delta$ produces one which sums exactly to a lattice point.
Hence a function with the original divisor is $F_\Delta$ times a product of
super theta functions as in Eq. (\ref{thetas}).
This completes the construction.

\section{The Picard and Jacobian Groups}

In this section we compute explicitly the Picard, Jacobian, and divisor class
groups of the super elliptic curve $M$.
These objects were defined and discussed in \cite{rsv}, where they were all
shown to be isomorphic in the normal case.
Some but not all of the arguments used there apply more generally;
nevertheless the isomorphisms will be verified here by direct calculation.
We also exhibit the Abel map from $M$ to its Jacobian, which is a projection
rather than an isomorphism as for classical elliptic curves.

We consider the set of line bundles over the superelliptic curve $M$.
A line bundle is specified by transition functions which are
elements of ${\cal O}^*_{\rm ev}$ \cite{giddings,rsv,rabinpt}, the
invertible, even functions, on overlaps of charts.
That is, the Picard group of line bundles under tensor product is
${\rm Pic}(M) = H^1(M,{\cal O}^*_{\rm ev})$ as usual.
The standard exponential exact sheaf sequence,
\begin{equation}
0 \rightarrow {\bf Z} \rightarrow {\cal O}_{\rm ev} \rightarrow
{\cal O}^*_{\rm ev} \rightarrow 1,
\end{equation}
and the resulting cohomology sequence,
\begin{equation}
H^1(M,{\bf Z}) \rightarrow H^1(M,{\cal O}_{\rm ev}) \rightarrow
H^1(M,{\cal O}^*_{\rm ev}) \rightarrow H^2(M,{\bf Z}),
\end{equation}
imply as usual that the group of line bundles of degree zero is
\begin{equation}
{\rm Pic}^0(M) = H^1(M,{\cal O}_{\rm ev})/H^1(M,{\bf Z}). \label{defpic}
\end{equation}

We can also describe a line bundle by the set of divisors of all its
meromorphic sections.
Since the ratio of two sections is a function, this gives an isomorphism
between ${\rm Pic}^0(M)$ and ${\rm Div}^0(M)$, the group of degree-zero
divisors modulo divisors of meromorphic functions \cite{giddings,rsv}.
We will compute both groups explicitly, verifying this isomorphism and
obtaining the projection map $\pi: M \rightarrow {\rm Pic}^0(M)$.

The divisor class group can be computed immediately from the results of the
previous section.
We first claim that every divisor $\Delta$ of degree zero is equivalent to one
of
the form $P-P_0$ with $P_0$ a fixed basepoint on $M$, for example $(0,0)$.
This is because $P$ can always be chosen so that $\Delta - P + P_0$ satisfies
the sum rules (\ref{sumrules}) and is therefore the divisor of a function.
What changes from the classical elliptic curve results is that the choice of
$P$ is not unique: evidently we are free to add multiples of $\delta$ to the
even coordinate of $P$ without changing the equivalence class of the divisor
$P-P_0$.
This establishes the central result of this section: the Abel map
$\pi: M \rightarrow {\rm Div}^0(M)$ which sends a point $P$ to the divisor
class $[P-P_0]$ is a projection onto
${\rm Div}^0(M) \cong M/ \!\sim$, where the identification is
$(z,\theta) \sim (z+\alpha\delta,\theta)$.
In the split case $\delta=0$ we recover the naive isomorphism of $M$ with
${\rm Div}^0(M)$ which might have been expected.

Before we confirm this result by direct computation of the Picard group, let us
pause to explain in the context of the group structure why $M$ cannot be
isomorphic to its Picard group in general.
The set of line bundles obviously carries the Abelian group structure given by
tensor product.
However, $M$ carries no such group structure.
Recall that $M$ is the quotient of ${\bf C}^{1,1}$ by the nonabelian group $G$.
Now, ${\bf C}^{1,1}$ itself can be identified with the nonabelian
supertranslation group,
\begin{equation}
(z,\theta) \cdot (w,\chi) = (z+w+\theta\chi,\theta+\chi).
\end{equation}
$G$ is the discrete subgroup generated by $(1,0)$ and $(\tau,\delta)$ acting by
right multiplication.
In view of the fact that \cite{kanno}
\begin{equation}
(z,\theta) \cdot (\tau,\delta) \cdot (z,\theta)^{-1} = (\tau +
2\theta\delta,\delta),
\end{equation}
$G$ is not a normal subgroup and the quotient $M$ does not inherit the group
structure.
However, ${\bf C}^{1,1}$ also admits an Abelian group structure via
\begin{equation}
(z,\theta) + (w,\chi) = (z+w,\theta+\chi).
\end{equation}
Of course, $M$ does not inherit this group structure either, because $G$ is not
a subgroup at all.
But let us take the quotient ${\bf C}^{1,1}/\!\sim$.
On this quotient space $G$ does act as a subgroup of the Abelian group
structure, hence a normal subgroup, and the further quotient by $G$ is the
Picard
group of $M$.
[Something is being swept under the rug here:
$\sim$ mods out by
all $\alpha\delta$ with $\alpha$ in the Grassmann algebra $\Lambda$.
This does not seem to include modding out by $\theta\delta$ as required to
identify $G$ as a subgroup.
One must remember that the group laws are really defined on the set of
$\Lambda$-valued points to resolve the apparent paradox.]
The moral is that the unexpected identification $\sim$ really provides the
minimal modification of $M$ which will admit an Abelian group structure as
${\rm Pic}^0(M)$ must.

We now turn to the direct computation of ${\rm Pic}^0(M)$ from (\ref{defpic}).
It seems cleanest to compute $H^1(M,{\cal O}_{\rm ev})$ as the group
cohomology $H^1(G,{\cal O}_{\rm ev})$ with values in the functions on
${\bf C}^{1,1}$, following similar calculations of Hodgkin
\cite{problems,direct}.
For an explanation of the equivalence between the sheaf
cohomology of $M$ and the group cohomology of $G$, see \cite{mumford}; the
techniques of group cohomology we use are fairly intuitive and can be found in
\cite[Appendix B]{silverman}.
In particular, there is the exact sequence
\begin{equation}
0 \rightarrow H^1[(S),\O^T] \rightarrow H^1(G,\O)
\rightarrow H^1[(T),\O],
\end{equation}
where $(T),(S)$ are the cyclic subgroups generated by the two generators of
$G$,
and $\O^T$ are the $T$-invariant functions.
The last cohomology group in this sequence is trivial, so we get the
isomorphism
\begin{equation}
H^1(G,\O) \cong H^1[(S),\O^T],
\end{equation}
which we use for our computation.
In geometric language this says that a torus is made from the plane by first
making the cylinder with fundamental group $(T)$, whose sheaf cohomology is
trivial because it is noncompact.
The cohomology of the torus is then computed directly from functions $\O^T$ on
the cylinder by identifying its ends with $S$.

A cocycle for $H^1[(S),\O^T]$ is determined by assigning to the generator $S$
a $T$-invariant function $F=f(z)+\theta\phi(z)$; it is trivial (exact) if
$F = \tilde{F} - S\tilde{F}$ for some $T$-invariant function
$\tilde{F}=g(z)+\theta\gamma(z)$.
This requires
\begin{equation}
f(z) + \theta \phi(z) = g(z) + \theta\gamma(z) - g(z+\tau+\theta\delta)
- (\theta+\delta)\gamma(z+\tau+\theta\delta),
\end{equation}
which amounts to
\begin{eqnarray}
f(z) & = & g(z) - g(z+\tau) - \delta\gamma(z+\tau),  \nonumber \\
\phi(z) & = & \gamma(z) - \gamma(z+\tau) - \delta g'(z+\tau).
\end{eqnarray}
Because every function appearing here is $T$-invariant, which is to say
periodic, they have Fourier series expansions of the form,
\begin{equation}
f(z) = \sum_{n=-\infty}^{\infty} f_n e^{2 \pi i n z},
\end{equation}
and similarly for the other functions.
Then the triviality of the cocycle becomes the conditions on the Fourier
coefficients,
\begin{eqnarray}
f_n & = & g_n(1-e^{2 \pi i n \tau}) - \delta \gamma_n e^{2 \pi i n \tau},
\nonumber \\
\phi_n & = & \gamma_n(1-e^{2 \pi i n \tau}) - 2\pi in\delta g_n e^{2 \pi i n
\tau}.
\end{eqnarray}
Given $f_n$ and $\phi_n$, these equations can always be solved for $g_n$ and
$\gamma_n$, {\it except} in the case $n=0$ when the conditions for triviality
are
\begin{equation}
f_0 = -\delta \gamma_0, \;\;\;\; \phi_0 = 0.
\end{equation}
That is, the nontrivial cocycles are precisely the odd constants and the even
constants modulo multiples of $\delta$: $H^1(M,\O) = {\bf C}^{1,1}/\!\sim$.

To complete the calculation, we must compute $H^1(G,{\bf Z})$.
Of course this is a lattice ${\bf Z} \oplus {\bf Z}$, but we need to know where
this lattice sits inside $H^1(G,\O)$.
An element of $H^1(G,{\bf Z})$ assigns integers $-n,m$ to the generators $T,S$
respectively.
In the calculation above, however, we used the triviality of $H^1[(T),\O]$ to
represent each class in $H^1(G,\O)$ by a cocycle which assigned zero to the
generator $T$.
To find such a representative of our element of $H^1(G,{\bf Z})$, we pick a
function $g(z)$ such that $-n = g(z) - g(z+1)$, for example $g(z)=nz$, and
subtract the trivial cocycle which assigns
\begin{eqnarray}
T & \mapsto & g(z) - g(z+1) = -n, \nonumber \\
S & \mapsto & g(z) - g(z+\tau+\theta\delta) = -n\tau - n\theta\delta,
\end{eqnarray}
obtaining the new representative
\begin{equation}
T \mapsto 0, \;\;\;\; S \mapsto m + n\tau + n \theta\delta.
\end{equation}
In terms of our identification $H^1(M,\O) = {\bf C}^{1,1}/\!\sim$, the elements
of $H^1(M,{\bf Z})$ are thus precisely the lattice points
$m(1,0)+n(\tau,\delta)$
in ${\bf C}^{1,1}$.
This explicitly shows that
\begin{equation}
{\rm Pic}^0(M) = H^1(M,\O)/H^1(M,{\bf Z}) = M/\!\sim \;\;= {\rm Div}^0(M).
\end{equation}

Next we wish to similarly calculate the Jacobian of $M$, defined \cite{rsv} as
the set of odd ($\Lambda$-)linear functionals on the holomorphic differentials
of
weight $1/2$, modulo those functionals which are the periods of the
differentials around cycles.
A $1/2$-differential on a super Riemann surface is a section of the canonical
bundle, the bundle whose transition functions are the Berezinian determinants
of those of $M$.
Since supertranslations (\ref{group}) have unit determinant, this bundle is
trivial for superelliptic curves, and the $1/2$-differentials can be identified
with functions.
The periods of such a function are obtained by integrating it over all homology
cycles.
Equivalently, we can lift a function $F$ to the covering space ${\bf C}^{1,1}$
and find an antiderivative $\Phi$ with $F=D\Phi$; the periods are the changes
in
$\Phi$ under the covering transformations generated by $T$ and $S$.
The Jacobian is then the set of odd linear functionals on
$H^0(M,{\cal O}) = \{a + \theta\alpha: \alpha\delta=0\}$ modulo periods.
Note that we consider all global functions, not merely even ones, so as to
obtain a $\Lambda$-module rather than a $\Lambda_{\rm ev}$-module.

The periods of the function $a + \theta \alpha$ are easily found.
An antiderivative is $\Phi = \alpha z + \theta a$.
Under the translation $T$ this changes by $\alpha$, while under the other
generator $S$ it changes by $\alpha\tau + \delta a$.
The odd linear functionals which send $a + \theta\alpha$ to integral linear
combinations of these two constants will be equivalent to zero in the
Jacobian.

To understand the structure of the linear functionals on the functions
$a+\theta\alpha$ let us begin with the simpler case in which $\delta$ is one
of the generators of the Grassmann algebra $\Lambda$.
Then the set of $\alpha$ which annihilate $\delta$ is just the set
of multiples of $\delta$, and a function $a + \theta \alpha$ is a linear
combination of the functions $1$ and $\theta\delta$.
Then an odd linear functional is determined by sending $1$ to some odd constant
$\eta$, and sending $\theta\delta$ to some odd constant $\kappa$.
By linearity, $\delta\kappa=0$, so $\kappa=\delta k$ for an even constant $k$
defined modulo $\delta$.
Hence we have found that the odd linear functionals correspond precisely to
points $(k,\eta)$ in ${\bf C}^{1,1}/\!\sim$.
They can be viewed as mapping $1 \mapsto \eta$ and $\theta \mapsto k$, just as
if $1$ and $\theta$ formed a basis for the functions, except that $k$ is only
defined modulo $\delta$.
Since the periods are just the familiar lattice points
generated by $(1,0)$ and   $(\tau,\delta)$, we  have explicit agreement between
the Jacobian and the Picard group computed eariler.
One can easily verify that
the isomorphism between them is the one described in \cite{rsv}: given a line
bundle in ${\rm Pic}^0$, represent it by a divisor in the form $P-P_0 =
(k,\eta)-(0,0)$ and associate to it the linear functional which integrates a
function from $P_0$ to $P$, which will also be $(k,\eta)$ with our conventions.

What changes in the general case in which $\delta$ is not a generator of
$\Lambda$?
A linear functional is still determined by its effect on the functions of the
forms $a$ and $\theta\alpha$ separately.
A functional on $\{a\}$ is still determined by the odd constant $\eta$ which is
the image of $1$, but the functionals on $\{\theta\alpha\}$ are not so clear.
We are asking for the $\Lambda$-linear functionals on the ideal
$I = \{\alpha: \alpha\delta=0\}$, the annihilator of $\delta$.
Because $\Lambda$ is an example of a quasi-Frobenius, or self-injective ring
\cite{small}, any such functional is multiplication by an even constant $k$
\cite{cartan} which is determined up to constants annihilating $I$.
Again because $\Lambda$ is self-injective, these are the multiples of $\delta$
\cite{dieudonne}.
Hence the isomorphism of the Picard and Jacobian groups holds in general.
To see that $\Lambda$ is indeed self-injective one can apply a simple
test from \cite{dieudonne}: the annihilator of the annihilator of any minimal
ideal of $\Lambda$ must be the ideal itself.
The unique minimal ideal in the Grassmann algebra with generators
$\beta_1,\beta_2,\ldots,\beta_N$ is the set of multiples of
$\beta_1\beta_2\cdots\beta_N$; its annihilator is the ideal of all nilpotents,
whose annihilator is indeed the minimal ideal again.

\section{The Group Law in a Projective Embedding}
As shown in \cite{rabinfreund}, the superelliptic curve $M$ can be embedded in
the projective superspace $P^{3,2}$ with the help of the super Weierstrass
function $R(z,\theta)$.
Indeed, the map
\begin{equation}
(z,\theta) \mapsto (R,R',R'',1;DR,D^3R) = (x,y,u,v;\phi,\psi)
\end{equation}
in the affine chart $v=1$, with the extension to the points at infinity,
\begin{equation}
(0,\theta) \mapsto (0,0,1,0;0,\theta),
\end{equation}
embeds $M$ as the locus of points satisfying the following homogeneous
polynomial equations:
\begin{eqnarray}
y^2v - 4x^3 + g_2xv^2 + g_3v^3 - 2 \phi \psi v & = & 0,  \nonumber \\
2y\psi v + (g_2v^2 - 12x^2)\phi + \delta\dot{g}_2xv^2 +
\delta \dot{g}_3v^3 & = & 0,  \nonumber \\
2yuv +(g_2v^2 - 12x^2)y - \delta \dot{g}_2\phi v^2 & = & 0, \nonumber \\
2(g_2v^2 - 12x^2)uv + (g_2v^2 - 12x^2)^2 + 2\delta \dot{g}_2 \psi v^3 & = & 0,
\label{polys}
\end{eqnarray}
where $g_2(\tau)$ and $g_3(\tau)$ are the usual modular functions.
The last equation is redundant except when $y=0$; $M$ is not a complete
intersection.

Now although $M$ is a variety, it does not carry a group structure; its
Jacobian, which does, is not a variety since varieties cannot have the kind of
singularities produced by the identification $\sim$: the reduced space is a
smooth manifold but not every $f(z)$ is a function on $M/\!\sim$ even locally
\cite{howdiff}.
What then becomes of the standard geometric implementation of
the group law by intersecting an elliptic curve with lines?

We attempt to follow the usual construction by taking a meromorphic function
$F$ on $M$ given by
\begin{equation}
F = aR + R' + \alpha DR + \beta D^3R + b.
\end{equation}
This is the restriction to $M$ of a linear function on $P^{3,2}$ (in the chart
$v=1$),
\begin{equation}
F = ax + y + \alpha \phi + \beta \psi + bv.
\end{equation}
The conditions for $F$ to have a principal zero at some point on $M$,
$F = DF = 0$, translate into the linear equations of a plane,
\begin{eqnarray}
ax + y + \alpha \phi + \beta \psi + bv & = & 0, \nonumber  \\
a\phi + \psi - \alpha y - \beta u & = & 0,  \label{plane}
\end{eqnarray}
to be solved simultaneously with the equations of $M$.
Note that this is hardly a generic plane, but rather a very special one
encoding the notion of a principal zero.
It is given by simple linear equations only because the embedding of $M$ was
constructed using the covariant derivative $D$ which also encodes the
superconformal structure.
We can adjust the four parameters $a,b,\alpha,\beta$ so that $F$ has principal
simple zeros at any two given points $P_i = (z_i,\theta_i), i = 1,2$ on $M$.
The naive expectation would be that $F$ has a principal triple pole at $(0,0)$
and, as a consequence of our function theory, there is a third point of
intersection with $M$ at $P_3$ such that $P_1 + P_2 + P_3 = 0\;{\rm mod} \sim$.
This turns out to be wrong on two counts.
First, using the fact that the singular part of $R(z,\theta)$ is $1/z^2$, we
find for the singular part of $F$
\begin{eqnarray}
F & \sim & az^{-2} - 2z^{-3} - 2 \alpha \theta z^{-3} + 6 \beta \theta z^{-4}
\nonumber \\
& = & (az - 2 - 3a\theta\beta + 2\theta\alpha)(z - \theta\beta)^{-3},
\end{eqnarray}
so that the triple pole is actually located at $(0,\beta)$.
This is a consequence of the fact that the most singular term in $F$ is the
nilpotent $\beta D^3R$ term.
We could not have avoided this by including an equally singular even term $R''$
in $F$, since then the condition $DF=0$ for a principal zero would involve
$D^5R$, which is not one of the projective coordinates in our embedding.
Next, there will indeed be a third point of intersection, another simple zero
of $F$ at $P_3$, but there is also a fourth intersection at the location of the
triple pole itself: $(0,\beta)$ embeds in $P^{3,2}$ as $(0,0,1,0;0,\beta)$,
which is easily seen to satisfy the homogeneous equations (\ref{plane}).
Thus the group law is realized in the form
\begin{equation}
P_1 + P_2 + P_3 - 3(0,\beta) = 0 \; {\rm mod}  \sim.
\end{equation}
This is a translate of the standard group law, with the identity shifted to the
point $(0,3\beta)$ in the fiber of $M$ at infinity.
Note that the point which plays the role of the identity varies with the choice
of points $P_1,P_2$ to be added, since $\beta$ depends on this choice, but it
can always be located geometrically as the fourth intersection of the curve
with the plane.
The existence of this fourth intersection could have been expected from the
fact the the reduction of this embedding of $M$ is not the usual degree 3
embedding of an elliptic curve in $P^2$, but the degree 4 embedding in $P^3$
using $\wp,\wp',$ and $\wp''$, in which there is indeed an extra
intersection at infinity \cite{weil}.

\section{Isogenies}
An isogeny of elliptic curves is a holomorphic map $f$ from one to the other
with
the translation symmetry normalized out by requiring $f(0)=0$.
One proves that an isogeny is either constant or onto, and that it is always a
homomorphism of the group structures.
Since a super elliptic curve does not have a group structure, the super
generalization will be that an isogeny ${\bf F}$ induces a group homomorphism
via
the projection maps to ${\rm Pic}^0$:
\begin{equation}
{\rm Pic}^0(M_1) \stackrel{\pi_1^{-1}}{\rightarrow} M_1
\stackrel{\bf F}{\rightarrow} M_2 \stackrel{\pi_2}{\rightarrow} {\rm
Pic}^0(M_2).
\label{isodiag}
\end{equation}
The homomorphism is independent of the inverse chosen for $\pi_1$.
We will also discuss isogenies from a super elliptic curve to itself and show
that only a split curve can admit nontrivial endomorphisms.
This is due to a conflict between the linear nature of an isogeny and the
quadratic constraint which is implicit in the superconformal structure of $M$.

Given two superelliptic curves $M_i = {\bf C}^{1,1}/G_i$ over $\Lambda$, with
$G_i$ generated by supertranslations of the form (\ref{group}) with parameters
$\tau_i,\delta_i$, an isogeny will be a holomorphic map ${\bf F}:M_1
\rightarrow M_2$ with  ${\bf F}(0,0) = (0,0)$.
(We will eventually require the map to be surjective as well.)
Its lift to the covering space ${\bf C}^{1,1}$ takes the form,
\begin{equation}
(z,\theta) \mapsto {\bf F}(z,\theta) = [F(z,\theta),\Psi(z,\theta)] =
[f(z) + \theta \phi(z), \psi(z) + \theta g(z)],  \label{isogeny}
\end{equation}
with $f(0) = \psi(0) = 0$.
Note that an isogeny is not assumed to be superconformal, but merely
holomorphic, even though the groups $G_i$ act superconformally.

In order that the map (\ref{isogeny}) descend to the quotient spaces $M_i$, it
is necessary and sufficient that acting on $(z,\theta)$ with a generator of
$G_1$
must change ${\bf F}(z,\theta)$ by the action of some element of $G_2$, which
must be independent of $z$ by continuity and the discreteness of the group.
Therefore, we have
\begin{eqnarray}
& & F(z+1,\theta) = F(z,\theta) + k + l\tau_2 + l\Psi(z,\theta) \delta_2, \\
& & \Psi(z+1,\theta) = \Psi(z,\theta) + l \delta_2, \\
& & F(z+\tau_1+\theta\delta_1,\theta+\delta_1) = F(z,\theta) + m + n\tau_2 +
n\Psi(z,\theta)\delta_2, \\
& & \Psi(z+\tau_1+\theta\delta_1,\theta+\delta_1) = \Psi(z,\theta) + n\delta_2,
\end{eqnarray}
with integers $k,l,m,n$.
If we use (\ref{isogeny}) to write these conditions in terms of
$f,\phi,\psi,g$, we obtain
\begin{eqnarray}
& & f(z+1) - f(z) = k + l\tau_2 + l\psi(z)\delta_2, \label{iso1} \\
& & \phi(z+1) - \phi(z) = lg(z)\delta_2, \label{iso2} \\
& & \psi(z+1) - \psi(z) = l\delta_2, \label{iso3} \\
& & g(z+1) - g(z) = 0, \label{iso4} \\
& & f(z+\tau_1) - f(z) = m + n\tau_2 + n\psi(z)\delta_2 -
\delta_1 \phi(z+\tau_1), \label{iso5} \\
& & \phi(z+\tau_1) - \phi(z) = ng(z)\delta_2 - \delta_1 f'(z+\tau_1),
\label{iso6} \\
& & \psi(z+\tau_1) - \psi(z) = n\delta_2 - \delta_1 g(z+\tau_1),
\label{iso7} \\
& & g(z+\tau_1) - g(z) = -\delta_1 \psi'(z+\tau_1). \label{iso8}
\end{eqnarray}

The analysis of these equations is somewhat tedious, but straightforward.
Eqs. (\ref{iso4}) and (\ref{iso8}) imply that $\delta_1g(z)$ is an elliptic
function, and entire, hence a constant.
(A simple argument using the filtration of $\Lambda$ shows that this is true
even though $\tau_1$ may have a nilpotent part.)
Given this, Eqs. (\ref{iso3}) and
(\ref{iso7}) say that $\psi'(z)$ is elliptic, hence constant.
Calling the constant $\gamma$ and using the normalization $\psi(0)=0$, we have
$\psi(z) = \gamma z$.
According to (\ref{iso3}), $\gamma = l\delta_2$.
{}From (\ref{iso7}),
\begin{equation}
\delta_1 g(z) = n\delta_2 - \gamma \tau_1 = (n-l\tau_1)\delta_2,
\end{equation}
so that $\delta_2$ must be a multiple of $\delta_1$ [and vice versa if we
assume
$g(z)$ is invertible].
Consequently, multiplying any equation by $\delta_1$ will kill terms containing
either $\delta_i$, and terms involving $\psi(z)\delta_i$ are already zero.

With this information, Eqs. (\ref{iso2}) and (\ref{iso6}) say that $\delta_1
\phi(z)$ is elliptic, so constant.
Then (\ref{iso1}) and (\ref{iso5}) say that $f'(z)$ is elliptic, which together
with the normalization $f(0)=0$ gives $f(z) = az$ where the constant
$a=k+l\tau_2$.
Eqs. (\ref{iso4}) and (\ref{iso8}) give that $g(z)$ is elliptic; so
$g(z) = c$, a constant, and (\ref{iso2}) and (\ref{iso6}) make $\phi'(z)$ a
constant, so $\phi(z) = \alpha z + \beta$ with $\alpha = lc\delta_2$ according
to
(\ref{iso2}).
Having expressed all the unknown functions in terms of a few
constants, all eight equations are satisfied provided the constants satisfy a
few relations. Eq. (\ref{iso5}) requires $\delta_1\beta = m+n\tau_2
-a\tau_1$;
Eq. (\ref{iso6}) gives
$\delta_1 a = nc\delta_2 - \alpha \tau_1 = (n-l\tau_1)c\delta_2$; and Eq.
(\ref{iso7}) implies
$\delta_1 c = n\delta_2 - \gamma \tau_1 = (n-l\tau_1)\delta_2$.
Collecting all these results, the general form of an isogeny is given by
\begin{eqnarray}
& & f(z) = az, \;\;\;\; \phi(z) = \alpha z + \beta, \;\;\;\; \psi(z) = \gamma
z,
\;\;\;\; g(z) = c; \nonumber \\
& & (z,\theta) \mapsto [az+\theta(\alpha z + \beta), \; \gamma z + \theta c],
\label{iso}
\end{eqnarray}
where
\begin{eqnarray}
& & a = k+l\tau_2, \;\;\;\; \gamma = l\delta_2, \;\;\;\; \alpha = c\gamma,
\nonumber \\
& &  \delta_1 a = (n-l\tau_1)c\delta_2, \;\;\;\; \delta_1c =
(n-l\tau_1)\delta_2, \;\;\;\; \delta_1 \beta = m + n\tau_2 - (k+l\tau_2)\tau_1
{}.
\label{isoconditions}
\end{eqnarray}

Having obtained this general form, we can use it to answer several questions
about isogenies of super elliptic curves.
First let us ask whether an isogeny, which is only holomorphic by definition,
is in fact a superconformal map.
A map ${\bf F}(z,\theta) = [F(z,\theta),\Psi(z,\theta)]$ is superconformal
provided that $DF = \Psi D \Psi$; in our case this says that
\begin{equation}
\alpha z + \beta + \theta a = \gamma cz + \theta c^2 .
\end{equation}
This requires $\alpha = \gamma c$, which is one of the conditions
(\ref{isoconditions}); $a=c^2$, which need only hold modulo the
annihilator of $\delta_1$ according to (\ref{isoconditions}); and $\beta=0$,
which is a completely new restriction.
We conclude that not every isogeny is superconformal; the superconformal ones
take the special form,
\begin{equation}
(z,\theta) \mapsto (c^2z + \theta\gamma cz, \gamma z + \theta c).
\end{equation}

Next, we see that while isogenies of ordinary elliptic curves are either
constant or onto, this is not true for super elliptic curves.
If the parameter $a$ is nilpotent, for example, a nonconstant isogeny may have
a
constant reduction, so that it is not surjective.
This is simply because the presence of nilpotents can lead to a wider range of
singularities for maps in general.
We prefer not to consider such singularities, so we assume from now on that all
our isogenies are surjective, which requires that the reduced parameters $a_\r$
and $c_\r$ be nonzero.
The important consequence of this is that $\delta_1$ is a multiple of
$\delta_2$ as well as vice-versa.

We now prove that a surjective isogeny of super elliptic curves induces a
well-defined homomorphism of their Picard groups via the diagram
(\ref{isodiag}), \begin{equation}
{\rm Pic}^0(M_1) \stackrel{\pi_1^{-1}}{\rightarrow} M_1
\stackrel{\bf F}{\rightarrow} M_2 \stackrel{\pi_2}{\rightarrow} {\rm
Pic}^0(M_2).
\end{equation}
A point $(z,\theta)$ of ${\rm Pic}^0(M_1)$ is the image under $\pi_1$ of any
point $(z+\epsilon\delta_1,\theta)$ of $M_1$ for any $\epsilon$.
The isogeny $\bf F$ sends this point to
\begin{equation}
(z+\epsilon\delta_1,\theta) \mapsto
[az + a\epsilon\delta_1 + \theta(\alpha z + \beta) +
\theta\alpha\epsilon\delta_1,
\gamma z + \gamma\epsilon\delta_1 + \theta c]
\end{equation}
in $M_2$.
Then $\pi_2$ removes any multiple of $\delta_2$ from the first coordinate.
The result is indeed independent of $\epsilon$, showing that the composite map
is well-defined, because the surjectivity makes $\delta_1$ a multiple of
$\delta_2$.
This also eliminates the term $\gamma\epsilon\delta_1$ from the second
coordinate, because the conditions (\ref{isoconditions}) include
$\gamma=l\delta_2$.

Now, at the level of the Picard groups, we can drop $\alpha$, which
is a multiple of $\delta_2$, from (\ref{iso}) and write an isogeny as
\begin{equation}
(z,\theta) \mapsto (az+\theta\beta, \gamma z + \theta c).
\end{equation}
But this is a linear map, and the group law is simply addition in these
coordinates, so the map is a group homomorphism as
claimed.

Next we examine isogenies of a super elliptic curve $M$ onto itself
(endomorphisms).
Setting $\tau_1 = \tau_2 = \tau, \;\; \delta_1 = \delta_2 = \delta$
in the general formulas, we obtain in this case
\begin{equation}
(z,\theta) \mapsto [az+\theta(\alpha z + \beta), \gamma z + \theta c],
\end{equation}
with
\begin{eqnarray}
& & \delta c = \delta (n-l\tau) ,\;\;\;\; \delta a = \delta c^2 ,
\label{modann}   \\
& & a=k+l\tau,\;\;\;\; \gamma=l\delta,\;\;\;\;
\alpha=cl\delta,\;\;\;\; \delta\beta=m+n\tau -(k+l\tau)\tau.
\label{endconditions} \end{eqnarray}
In the special case when $M$ is split, $\delta=0$, we lose the conditions
(\ref{modann}) and obtain the simple form,
\begin{eqnarray}
& & (z,\theta) \mapsto (az + \theta\beta, \theta c), \\
& & a = k+l\tau,\;\;\;\; 0 = m+n\tau - (k+l\tau)\tau .
\end{eqnarray}
In particular, $c$ is now arbitrary; there is no relation like $a=c^2$ in this
case.

We see that in the split case, multiplication by an integer $k$,
$(z,\theta) \mapsto (kz,k\theta)$, is an endomorphism, which was to be expected
since $M$ and its Picard group coincide in this case.
But this is not true more generally, since this map violates the condition
$\delta a = \delta c^2$ which is a vestige of the superconformal action of the
group $G$.
In fact, for $\delta \neq 0$, this implies $a_\r = c_\r^2$, which gives the
quadratic constraint,
\begin{equation}
l^2\tau_\r^2 - (2n+1)l\tau_\r + (n^2 - k) = 0.
\end{equation}
This must hold in addition to the usual quadratic constraint appearing in the
theory of complex multiplication, which here arises from reducing the condition
on $\delta\beta$ in (\ref{endconditions}),
\begin{equation}
l\tau_\r^2 + (k-n)\tau_\r - m = 0.
\end{equation}
When $l \neq 0$ we are indeed describing complex multiplication, meaning an
endomorphism with $a$ complex.
By eliminating the quadratic term between these equations, we conclude
that $\tau_\r$ is rational, not complex, a contradiction which shows that a
nonsplit $M$ cannot admit complex multiplication.
However, even in the case $l=0$ when $a$ is an integer, the constraints give
$k=n^2$ in addition to the usual $k=n$ and $m=0$, so that $M$ admits only the
trivial endomorphisms $k=n=0,1$.

\section{Supercommuting Differential Operators from Super \protect\\ Elliptic
Curves}

The beautiful Krichever theory which produces solutions to the
Kadomtsev--Petviashvili (KP) hierarchy of nonlinear PDEs from geometric data
consisting of a line bundle over an algebraic curve together with some
coordinate choices is by now well-known \cite{segal,mulasenotes}.
The simplest explicit example uses a line bundle $\cal L$ of degree zero over
an
elliptic curve $M$ to construct the commuting pair of ordinary differential
operators,
\begin{eqnarray}
Q & = & \frac{d^2}{dx^2} - 2 \wp(x+a), \\
P & = & Q_+^{3/2} = \frac{d^3}{dx^3} - 3 \wp(x+a)\frac{d}{dx} - \frac{3}{2}
\wp'(x+a),
\end{eqnarray}
where $Q_+^{3/2}$ is the differential operator part of $Q^{3/2}$
computed in the larger algebra of formal pseudo\-differential operators.
The correspondence which associates $Q$ and $P$ to the
meromorphic functions $\wp(z)$ and $-\wp'(z)/2$ on $M$ respectively sets up an
isomorphism between the commutative ring of differential operators generated by
$Q,P$ and the ring of meromorphic functions on $M$ with poles only at $z=0$,
which is generated by $\wp(z)$ and $-\wp'(z)/2$.
As $\cal L$ varies through the Picard group ${\rm Pic}^0(M)$, the parameter $a$
changes and the ring of operators is isospectrally deformed.
In fact, there is an infinite set of linear coordinates $t_n$ for
${\rm Pic}^0(M)$ on which $a$ depends linearly, with $Q$ satisfying the KP
equations,
\begin{equation}
\frac{\partial Q}{\partial t_n} = [Q_+^{n/2},Q].
\end{equation}

The corresponding construction of solutions to the supersymmetric KP
hierarchies was worked out recently \cite{rabin,mulase,supercomm}.
One surprise was that the geometric data involve a line bundle over a specific
type of algebraic supercurve, which cannot be a super Riemann surface except in
the special case of genus one.
Another was the fact that linear flow in the Picard group of a fixed supercurve
is described by a new super KP hierarchy discovered by Mulase and myself, and
not by either of the previously known hierarchies due to Manin--Radul or to
Kac--van de Leur.
It follows that explicit solutions to this new super KP hierarchy can be
constructed using the information about the Picard group of a super elliptic
curve developed in the previous sections.
In this section we exhibit and discuss these solutions.
We change our notation slightly to conform to the conventions of the literature
on KP theory: the standard coordinates on the covering space ${\bf C}^{1,1}$ of
the supertorus $M$ will now be denoted by $(w,\phi)$, so that $(z,\theta)$ can
be
reserved for a different set of local coordinates on $M$ to be introduced
below.

We begin with an overview of the construction to be carried out.
In a small disk $U$ around the point $P_0: (w,\phi) = (0,0)$ we introduce new
coordinates $(z,\theta)$ such that $z^{-2}$ and $\theta z^{-3}$ extend to
global holomorphic functions on $M-P_0$.
We fix a nontrivial line bundle $\cal L$ of degree zero on $M$ and note that it
is holomorphically trivial on each of the Stein patches $U$ and $M-U$, hence
completely described by a transition function across the overlap, a small
annular neighborhood of $\partial U$, which we can take to be the circle $z=1$.
We embed $\cal L$ in a family of bundles ${\cal L}(x,\xi)$ by multiplying its
transition function by an extra factor $\exp (xz^{-1} + \xi\theta)$.
Although these bundles have no holomorphic
sections, they have one which has the form ($z^{-1} +$ holomorphic) near $P_0$
(note that this is different from having a principal simple pole there); the
expression of this section in the coordinates $(z,\theta)$ in the chart $M-U$
is
the Baker-Akhiezer function $B(z,\theta,x,\xi)$ [although we will express it
in terms of the covering space coordinates $(w,\phi)$ instead].
It is the basic object in the theory and we will construct it explicitly in
terms
of super theta functions.
We observe that successive derivatives of $B$ with respect to $x$ and
$\xi$ produce sections having poles of higher orders at $P_0$ and constitute a
basis for the space of meromorphic sections on $M$ with poles only at $P_0$.
This allows us to set up an isomorphism between the ring of functions having
poles only at $P_0$ and a ring of super differential operators as follows.
Given such a meromorphic function $F$, $FB$ is a
section with poles at $P_0$ only, so it must be a linear combination of
derivatives of $B$.
But this is to say that it arises from $B$ by the action of
a certain differential operator $O_F$, so we associate this operator to $F$.
It can be computed for an explicit $F$ by matching the singular and
constant terms in the Laurent expansions of $FB$ and $O_FB$ about $P_0$.
We will exhibit a set of
generators for this ring analogous to $Q,P$ above, and discuss how they flow
under the deformations of $\cal L$ described by the super KP equations.

We start with the specification of the new coordinates $(z,\theta)$.
In order that $z^{-2}$ extend to a holomorphic function away from $P_0$, we
choose
\begin{equation}
z^{-2} = R(w,\phi) + \frac{2}{3}c = \wp(w;\tau+\phi\delta) + \frac{2}{3}c,
\end{equation}
where $c$ is a constant and $R$ is the super Weierstrass function introduced
earlier.
Similarly, in order that $\theta z^{-3}$ extend holomorphically we use a
function with behavior $\phi w^{-3}$ near $P_0$, setting
\begin{equation}
-2\theta z^{-3} = DR(w,\phi) - 2\gamma =
\delta \dot{\wp}(w;\tau) + \phi \wp'(w;\tau) - 2\gamma,
\end{equation}
where $\gamma$ is another constant, and we recall that a dot means
$\partial_\tau$ while a prime denotes $\partial_w$.
Using the Laurent expansion
of $\wp(w)$ \cite{chandra} we obtain the relation between the two sets of
coordinates,
\begin{eqnarray}
& & z^{-1} = w^{-1}[1 + \frac{1}{3}cw^2 + (\frac{g_2}{40} -
\frac{c^2}{18} + \frac{\dot{g}_2}{40} \phi\delta) w^4 + \cdots ], \\
& & \theta = \phi - c\phi w^2 + \gamma w^3 +
(\frac{c^2}{2} - \frac{g_2}{8}) \phi w^4 -
(c\gamma + \frac{\dot{g}_2}{40}\delta) w^5 + \cdots .
\end{eqnarray}

Following the construction of the Baker function in the non-super theory
\cite{segal}, we express it as a ratio of super theta functions times a
prefactor which is the exponential of a function with the behavior
$xz^{-1} + \xi\theta = xw^{-1} + \xi\phi \;+$ holomorphic terms.
Such a prefactor is
\begin{equation}
\exp [x \partial_w \log H(w,\phi) + \xi\phi].
\end{equation}
It has the correct singular part because of
\begin{equation}
\partial_w \log H(w,\phi) = w^{-1} + (q + \phi\delta \dot{q})w + \cdots,
\end{equation}
where $q$ is still the ratio of theta constants introduced in (\ref{q}).
It is invariant under the covering transformation $T:w \rightarrow w+1,\;\;
\phi \rightarrow \phi$, while under the other generator $S$ it acquires a phase
\begin{equation}
\exp (-2\pi i x + \xi\delta) = \exp -2\pi i (x-\frac{\xi\delta}{2\pi i}).
\label{expphase}
\end{equation}
As our ``pre-Baker function'' $\hat{B}$ we take the product of this with a
ratio
of theta functions transforming by the opposite phase, namely
\begin{equation}
\hat{B} = \exp [x\partial_w \log H(w,\phi) + \xi\phi]
\frac{H(w-a-\phi\alpha-x+\frac{\xi\delta}{2\pi i}, \phi + \alpha)}{
H(w-a-\phi\alpha, \phi + \alpha)},  \label{preBaker}
\end{equation}
as is easily verified using (\ref{Sphase}).

The parameters $a$ and $\alpha$ describe the given line bundle $\cal L$: its
divisor is $(a,\alpha) - (0,0)$.
It has a section given by $1$ outside the disk $U$, and
$z^{-1}H(w-a-\phi\alpha, \phi + \alpha)$ inside.
Equivalently, its transition function across $\partial U$ (inside to outside)
is
$z/H(w-a-\phi\alpha, \phi + \alpha)$.
Then the transition function of the deformed bundle ${\cal L}(x,\xi)$ is
\begin{equation}
\frac{z \exp (xz^{-1} + \xi\theta)}{H(w-a-\phi\alpha, \phi + \alpha)}.
\end{equation}
Now $\hat{B}$ is to be viewed as a section of this bundle in the outside chart
$M-U$; dividing by the transition function gives the same section in the inside
chart $U$ as a nonvanishing holomorphic function (the mismatch between the
exponential factors) times
$z^{-1}H(w-a-\phi\alpha-x+\frac{\xi\delta}{2\pi i},\phi + \alpha)$, from which
we see that the deformed bundle has divisor
$(a+x-\frac{\xi\delta}{2\pi i},\alpha) - (0,0)$.
(We assume that all constants and parameters are small enough that the supports
of these divisors are inside $U$.)
In particular, $x$ shifts the even coordinate
of ${\rm Pic}^0(M)$ linearly and could be viewed as such a coordinate itself,
but
$\xi$ does {\it not} shift the odd coordinate $\alpha$.
In fact, because of the identification $\sim$, $\xi$ induces no flow on the
Picard group at all but only changes the trivialization of the bundle.

The pre-Baker function can be normalized so that, apart from the exponential
prefactor, its Taylor series in powers of $w$ and $\phi$ begins with constant
term unity.
We will need this series through the quadratic terms in order to match singular
parts of Laurent series later:
\begin{eqnarray}
\hat{B}_n & = &
\frac{\Theta(a;\tau+\alpha\delta)}
{\Theta(a+x-\frac{\xi\delta}{2\pi i};\tau+\alpha\delta)} \hat{B} \nonumber \\
& = & \exp [x \partial_w \log H(w,\phi) + \xi\phi]
\{ 1 + \phi\alpha L' + \phi\delta \dot{L} - wL'  \nonumber \\
& & - \phi\alpha w(L'' + L'^2) - \phi\delta w (\dot{L}' + L' \dot{L}) +
\frac{1}{2} w^2 (L'' + L'^2)   \nonumber \\
& &  +\frac{1}{2} \phi\alpha w^2 (L''' + 3L'L'' + L'^3) +
\frac{1}{2} \phi\delta w^2 (\dot{L}'' + 2L' \dot{L}' + L''\dot{L} +
\dot{L}L'^2) + \cdots \} / N,
\end{eqnarray}
where we have introduced the abbreviations
\begin{equation}
L = L(x,\xi,\tau,\delta) = \log \Theta(a+x-\frac{\xi\delta}
{2\pi i};\tau+\alpha\delta), \;\;\;\; L' = \partial_x L, \;\;\;\; \dot{L} =
\partial_\tau L,
\end{equation}
and the normalization constant $N$ is the series in braces with $x$ and $\xi$
set to zero.
Although the series has constant term unity, leading to the behavior $1/z$ for
this section near $P_0$, we see that there are also terms proportional to
$\phi$, leading to additional singularities like $\phi /z$.
To obtain the true Baker function, we must subtract these off.
Because of the exponential prefactor, derivatives of $\hat{B}_n$ with respect
to $x$ or $\xi$ produce new sections\footnote{
It may not be clear that
derivatives of $\hat{B}$ are still sections of ${\cal L}(x,\xi)$.
The point is that $\hat{B}$ is a global function on $M-U$
for all $x,\xi$, so its derivatives are too.
They must extend into $U$ as
meromorphic sections since no essential singularity has been introduced.}
containing additional factors $z^{-1}$ and $\phi$ respectively, so
$\partial_\xi
\hat{B}_n$ has a $\phi /z$ singularity.
Subtracting the appropriate multiple of
this yields the true Baker function,
\begin{eqnarray}
B & = & e^{[\cdots]} N^{-1} \{ 1 - wL' + \alpha\phi wL'' + \delta\phi w
\dot{L}'
+ \frac{\alpha\delta}{2\pi i} wL'L'' +
\frac{1}{2} w^2 (L'' + L'^2)  \nonumber \\
& & + \frac{1}{2} \phi\alpha w^2 (L''' + 2L'L'') + \frac{1}{2} \phi\delta w^2
(\dot{L}'' + 2L'\dot{L}') - \frac{\alpha\delta}{4\pi i} w^2 (L'L''' + 2L''L'^2)
+ \cdots \}.
\end{eqnarray}

It is now tedious but straightforward to work out the explicit correspondence
between meromorphic functions $F$ on $M$ holomorphic away from $P_0$ and
differential operators $O_F$ in $x,\xi$ by matching the singular terms in the
series for $FB = O_F B$.
For example, the operator corresponding to the super Weierstrass function
$R(w,\phi)$, with a double pole at $P_0$, has the form
$Q = d^2 + \omega \partial + u$, with
\begin{eqnarray}
\omega & = & 2[\alpha\wp'(a+x;\tau) + \frac{\alpha\delta\xi}{2\pi i}\wp'' +
\delta\dot{\wp}], \nonumber \\
u & = & 2\{-\wp + \frac{\xi\delta}{2\pi i}\wp' - \alpha\delta\dot{\wp} +
\alpha\delta\dot{q} + \frac{\alpha\delta}{2 \pi i}[\wp'\partial_x \log \Theta -
(\wp-q)^2]\}, \label{Q}
\end{eqnarray}
where all the functions have the same arguments as $\wp'(a+x;\tau)$, all
odd parameters having been explicitly expanded out, and $d = \partial_x,
\; \partial = \partial_\xi$.
It follows from the general theory, and can be verified explicitly, that the
function $-R'(w,\phi)/2$ having a triple pole must correspond to
$P = Q_+^{3/2}$.
For any such second-order operator $Q$, one finds
\begin{equation}
P = Q_+^{3/2} = d^3 + \frac{3}{2}\omega\partial d + \frac{3}{2}ud +
\frac{3}{4}\omega'\partial + \frac{3}{4}u'.
\end{equation}

A set of generators for the ring of functions holomorphic off $P_0$ must
contain an odd function in addition to $R,-R'/2$; this is conveniently taken to
be $\sigma(w,\phi)$ of Eq. (\ref{sigma}), which corresponds to the simple
first-order operator
\begin{equation}
\Sigma = \partial + \frac{\delta}{2\pi i}d.
\end{equation}
The supercommutativity of the generators $Q,P,\Sigma$ of the isomorphic ring of
operators can be verified explicitly.
Although it is not manifest from the form of (\ref{Q}), both $Q$ and $P$ depend
on $x,\xi$ only through the combination $x-\frac{\xi\delta}{2\pi i}$ [see
(\ref{expphase}),(\ref{preBaker}) for the origin of this], and this is
precisely
the statement that they commute with $\Sigma$.
We also have $\Sigma^2=0$.
The vanishing of $[Q,P]$ leads to a pair of third-order differential equations
for $\omega,u$, namely
\begin{eqnarray}
\omega_{xxx} + 3\omega\omega_{x\xi} + 6\omega u_x + 6u\omega_x +
3\omega_x\omega_{\xi} & = & 0, \\
u_{xxx} + 3\omega u_{x\xi} + 3\omega_x u_{\xi} + 6uu_x & = & 0.
\end{eqnarray}
One finds that, exactly as in the non-super case, the first equation is
satisfied identically in virtue of the identity
\begin{equation}
\wp''' = 12\wp\wp'
\end{equation}
satisfied by the Weierstrass function.
However, the second equation requires, in addition to this identity, the
relation
\begin{equation}
g_2 = 12(q^2 - 2\pi i \dot{q})  \label{relation}
\end{equation}
between the modular function $g_2$ and the theta constant $q$.
I have found similar relations in the literature on elliptic functions, though
not in just this form; however, it is a simple consequence of the fact that the
theta function satisfies the heat equation \cite{chandra},
\begin{equation}
4\pi i \dot{\Theta}(w;\tau) = \Theta''(w;\tau).
\end{equation}
As a consequence, its logarithm $f = \log \Theta$
satisfies
\begin{equation}
4\pi i \dot{f} = f'' + f'^2.  \label{diffeq}
\end{equation}
{}From the relation (\ref{q}) between $\Theta$ and $\wp$ we get the Laurent
expansion
\begin{equation}
f' = w^{-1} + qw - \frac{g_2}{60}w^3 + \cdots,
\end{equation}
and the desired relation (\ref{relation}) follows by using this in
(\ref{diffeq})
and equating the coefficients of $w^2$ on both sides.
This illustrates that the super KP system contains information about the
modular dependence of the theta functions, through the coupling between $\tau$
and $\theta$ in the superelliptic functions, which does not appear in the
solutions to ordinary KP (although changes in moduli do figure in the
additional symmetries of the KP hierarchy).

Finally, we describe the flows on the Picard group (further deformations of
$\cal L$) which lead to the super KP equations for $Q$.
These depend on an infinite set of parameters $t_n$ which are (Grassmann) even
or odd for even or odd $n$ respectively.
They multiply the transition function of ${\cal L}(x,\xi)$ by an additional
factor $\exp t_{2n}z^{-n}$ or $\exp t_{2n+1}\theta z^{-n}$ respectively.
At this point the properties of the new coordinates $(z,\theta)$ become
important.
Because $z^{-2}$ extends to a holomorphic function on $M-U$, all the flows
$t_4,t_8,\ldots$ are trivial since they can be undone by a change of bundle
trivialization on $M-U$.
Because $\theta z^{-3}$ extends holomorphically, the same is true for
$t_7,t_{11},\ldots$.
The parameter $t_2$ can be identified with $x$, since they produce the same
deformation.
The first nontrivial even flow is by $\exp t_6z^{-3}$, and we need to
understand the Baker function for the new bundle this produces.
It should have an exponential prefactor having this singular behavior.
For this purpose we employ the function $-R'(w,\phi)/2$ with singular part
$w^{-3} = z^{-3} - cz^{-1} + \cdots$.
Thus we need only multiply our previous Baker function by $\exp -t_6R'/2$ and
replace $x$ by $x+ct_6$ to obtain the new one.
The effect on the resulting differential operators is the replacement
$a \rightarrow a + ct_6$ showing explicitly the flow on the Jacobian where $a$
is the even coordinate.
The flow would be trivial if we had chosen $c=0$; the motivation for
introducing this constant is precisely to get a nontrivial $t_6$ flow.

Similarly, for the first nontrivial odd flow by $\exp t_3\theta z^{-1}$ an
exponential prefactor with this behavior is
\begin{equation}
\exp t_3 \partial_\eta \log H(w-\phi\eta, \phi + \eta) =
\exp t_3[\phi \frac{\Theta'(w;\tau)}{\Theta(w;\tau)} +
        \delta \frac{\dot{\Theta}(w;\tau)}{\Theta(w;\tau)}].
\end{equation}
This function is invariant under the generator $T$, but acquires a phase
\begin{equation}
\exp -t_3 (\pi i \delta + 2\pi i \phi)
\end{equation}
under $S$.
To obtain a well-defined pre-Baker function we compensate this phase by
shifting
the parameter $\alpha$ in the numerator factor
\begin{equation}
H(w-a-\phi\alpha-x+\frac{\xi\delta}{2\pi i},\phi+\alpha)
\end{equation}
in (\ref{preBaker}) by $\alpha \rightarrow \alpha - t_3$, which is the flow on
the Jacobian in this case.
The next odd flow $t_5\theta z^{-2}$ is actually trivial because there is a
global function with this behavior, namely
\begin{equation}
-t_5 D \partial_w \log H(w,\phi) = t_5 [\theta z^{-2} + \theta (\frac{c}{3}-q)
+ \cdots].
\end{equation}
The mechanics of this triviality is rather interesting: if this function is
used to form an exponential prefactor for the pre-Baker function, a shift of
the parameter $\xi$ will be required due to the term proportional to $\theta$.
We know that $\xi$ only changes the trivialization of a bundle, and indeed the
change in $\hat{B}$ resulting from this shift is subtracted off along with the
$\phi /z$ poles in forming $B$, so that the differential operators are
unchanged.

The higher flows can all be computed in the same manner.
Because there are global functions with leading singularities $z^{-n}$ and
$\theta z^{-n}$ for all $n \ge 2$, we can use them as prefactors for the Baker
function (that is, to change the bundle trivialization in $M-U$) until any flow
is reduced to a linear combination of those for $n=1$.
(In other words, any deformation can be reduced to a linear combination of the
single even and odd generators for ${\rm Pic}^0$.)
Then its effect can be read off as a linear shift in the Jacobian coordinates
$a$ and $\alpha$.
It is not always true, however, that the even flows only shift $a$ while the
odd
flows only shift $\alpha$.
In general each flow can shift both in the nonsplit situation.
The flow parametrized by $t_{10}$, for example, acts by
\begin{equation}
a \rightarrow a + (\frac{g_2}{8} + \frac{5}{6}c^2)t_{10},\;\;\;\;
\alpha \rightarrow \alpha - \frac{\dot{g}_2}{8} \delta t_{10},
\end{equation}
showing how the supermodulus $\delta$ permits a flow in both even and odd
coordinates.

The differential operator $Q+\frac{2}{3}c$ corresponding to the function
$z^{-2}$, with its parameters shifted in this manner, gives a solution to the
new
super KP hierarchy of \cite{rabin,mulase}.
Unfortunately, unlike the standard KP
theory, this hierarchy has no simple formulation in terms of $Q$ itself, but is
written in the Sato form in terms of the wave pseudodifferential operator $K$
which conjugates $Q$ into a simple form:
\begin{eqnarray}
Kd^2K^{-1} & = & Q+\frac{2}{3}c, \\
\frac{\partial K}{\partial_{t_{2n}}} & = & -(Kd^nK^{-1})_- K, \\
\frac{\partial K}{\partial_{t_{2n+1}}} & = & -(K\partial d^n K^{-1})_- K .
\end{eqnarray}
I have not tried to obtain an explicit expression for $K$.

\section{Conclusions}

In this paper we have developed the theory of super elliptic curves with an
emphasis on the role of the supermodulus $\delta$ and the non-freely generated
character of the cohomology modules.
We discussed the building blocks for superelliptic functions, the super
Weierstrass and theta functions, and proved the necessary and sufficient
conditions for a divisor to be the divisor of a superelliptic function.
We computed the Picard, Jacobian, and divisor class groups of a superelliptic
curve, explicitly verifying the isomorphisms between them, and found that the
Abel map $\pi: M \rightarrow {\rm Pic}^0(M)$ was a projection in the nonsplit
case. The agreement between the different methods of calculation --- cohomology
for the Picard group, duality of modules for the Jacobian, function theory for
the divisor class group --- is very satisfying.
We showed that the group law can be implemented in a projective embedding by
intersecting $M$ with planes chosen to encode the notion of principal zero,
modulo the kernel of $\pi$ and an ambiguity in the group identity element.
We determined the general form of an isogeny of superelliptic curves, proving
that it always induces a homomorphism of their Picard groups, and that a
nonsplit curve admits trivial endomorphisms only.
Finally, we applied this machinery to the explicit calculation of the
supercommutative rings of differential operators which constitute the solution
to the new super KP hierarchy corresponding to flow in the Jacobian
of a superelliptic curve.
The Baker function was expressed in terms of super theta functions and used to
work out the differential operators corresponding to simple superelliptic
functions, generalizing the classical $Q,P$ pair of ordinary KP theory.

It would be natural to seek extensions of this theory in two directions:
higher-genus super Riemann surfaces, and supercurves of genus one which are not
super Riemann surfaces.
For super Riemann surfaces of higher genus the primary motivation is again to
understand the consequences of the non-freely generated cohomology.
One should again construct the Picard, Jacobian, and divisor class groups as
explicitly as possible and check their isomorphism in the general nonsplit
case.
An Abel map from the surface to its Jacobian should be constructed and
investigated.
Function theory on the surface should be studied in terms of the pullback of
theta functions from the Jacobian.
A higher-genus analogue of the simple substitution
$\tau \rightarrow \tau + \theta\delta$ which converts ordinary theta functions
to super ones should be found.
As here, the duality properties of modules which determine the structure of the
Jacobian should be understandable on the basis of $\Lambda$ being
self-injective,
and this should be used to develop Serre duality for cohomology groups as
$\Lambda$-modules rather than as $\bf C$-modules as in \cite{wells}.
One should study the map from super Riemann surfaces with local
coordinates to states in the operator formalism, and the geometry of the
Grassmannian of such states when the ring of functions with poles at a single
point is not freely generated.

The motivation for studying genus-one supercurves which are not super Riemann
surfaces, or Abelian supergroups on two generators whose action on
${\bf C}^{1,1}$ need not be superconformal, is to construct more general
solutions to super KP hierarchies.
(One should also find nontrivial endomorphisms of such curves with the relaxing
of the superconformal constraint.)
We know from \cite{rabin} that the
Manin--Radul and Kac--van de Leur super KP hierarchies describe simultaneous
deformations of the supercurve $M$ and the line bundle $\cal L$ over it,
specifically by changing the patching of the coordinate $\theta$ along with
that of the line bundle across $\partial U$.
Even if $M$ is initially a super Riemann surface, this property will not be
preserved by the flow.
Hence one needs to repeat enough of the analysis of this
paper for general genus-one curves to construct the Baker functions for
families
of line bundles over such curves.
One may learn something about where the locus of super Riemann surfaces
sits inside the larger moduli space of genus-one curves by studying the
corresponding super KP solutions, e.g. what is special about the rings of
differential operators when $M$ admits a superconformal structure?
Without the covariant derivative $D$ one will have to settle for Cartier
divisors which are not sums of points.
On the other hand one may be able to exploit the remarkable correspondence
\cite{dolgikh} between general supercurves and untwisted $N=2$ super Riemann
surfaces, and the resulting involution in the moduli space under which $N=1$
super Riemann surfaces are fixed points.
Perhaps this involution plays a role in the super KP theory.

\section*{Appendix}

A natural question is whether anything of number-theoretic interest results
from seeking rational points on super elliptic curves.
By analyzing a simple example we will see that this essentially amounts to
finding rational points on the (co)tangent plane --- more generally, the jets
--- of an ordinary elliptic curve at a rational point.
This answers our question in the negative, since rational points on planes are
abundant and easy to find.

When we consider super elliptic curves over $\bf Q$, the generators of the
lattice cannot always be reduced to the form (\ref{group}).
Instead we must consider the more general form,
\begin{eqnarray}
& T: & z \rightarrow z + \omega_1 + \theta\delta_1, \;\;\; \theta \rightarrow
\theta + \delta_1, \nonumber \\
& S: & z \rightarrow z + \omega_2 + \theta \delta_2,
\;\;\; \theta \rightarrow \theta + \delta_2,
\end{eqnarray}
with $\delta_1\delta_2 = 0$.
As in \cite{rabinfreund}, we find that the affine part of the super elliptic
curve is embedded in ${\bf C}^{2,2}$ by the map,
\begin{equation}
(z,\theta) \mapsto (R,R';DR,D^3R) = (x,y;\phi,\psi),
\end{equation}
where $R(z,\theta) = \wp(z;\omega_1+\theta\delta_1,\omega_2+\theta\delta_2)$,
as the set of solutions of the polynomial equations,
\begin{eqnarray}
& & y^2 - 4x^3 + g_2x + g_3 - 2\phi\psi = 0, \nonumber  \\
& & 2y\psi - (12x^2-g_2)\phi + \sum_{i=1}^2 \delta_i(\partial_{\omega_i}g_2x +
\partial_{\omega_i}g_3) = 0.
\end{eqnarray}

We fix $\Lambda$ to be the
Grassmann algebra on just two generators $\beta_1,\beta_2$, and consider the
affine supertorus in ${\bf C}^{2,2}$ given by the equations,
\begin{eqnarray}
& & y^2 - 4x^3 + g_2x + g_3 - 2 \phi\psi = 0,  \nonumber \\
& & 2y\psi - (12x^2 - g_2)\phi + a\beta_1 x + b\beta_2 = 0, \label{poly}
\end{eqnarray}
where $g_2,g_3,a,b$ are rational.
Now $\Lambda$ is a four-dimensional vector space, and using the basis
$\{1,\beta_1,\beta_2,\beta_1\beta_2\}$ we can write
\begin{eqnarray}
& & x = x_\r + x_{12}\beta_1\beta_2,\;\;\;\; y = y_\r + y_{12}\beta_1\beta_2,
\nonumber \\
& & \phi = \phi_1\beta_1 + \phi_2\beta_2,\;\;\;\;
\psi = \psi_1\beta_1 + \psi_2\beta_2 .
\end{eqnarray}
By a rational point we understand one whose components in this basis are
rational.
Inserting these expressions into the polynomial equations (\ref{poly}), we
obtain
\begin{eqnarray}
& & y_\r^2 - 4x_\r^2 + g_2x_\r + g_3 = 0, \\
& & 2y_\r\psi_1 - (12x_\r^2 - g_2)\phi_1 = -ax_\r - b, \\
& & 2y_\r\psi_2 - (12x_\r^2 - g_2)\phi_2 = 0, \\
& & 2y_\r y_{12} - (12x_\r^2 - g_2)x_{12} = 2(\phi_1\psi_2 - \phi_2\psi_1).
\end{eqnarray}
The first equation says that $(x_\r,y_\r)$ must be a rational point on the
reduced elliptic curve.
The three remaining equations share a common structure which can be understood
by recalling the invariant differential of the reduced curve,
\begin{equation}
2y_\r dy_\r - (12x_\r^2 - g_2) dx_\r = 0.
\end{equation}
This can be viewed as defining a linear map from rational values of $dx_\r$ to
rational values of $dy_\r$, or vice versa, at the chosen rational point of
$M_\r$; the derivative map defined over the rationals.
Similarly here we get a map from rational $(\phi_1,\phi_2,x_{12})$,
playing the role of $dx_\r$, to rational $(\psi_1,\psi_2,y_{12})$ analogous to
$dy_\r$, which is a deformation of the derivative map and is computed by
solving linear equations only.
A Grassmann algebra with more generators will
bring higher derivatives into play through later terms in the Taylor expansions
of the equations (\ref{poly}).

\end{document}